\documentstyle[preprint,aps,epsf,graphicx]{revtex}
\begin{document}

%\tightenlines

\title{
The antikaon nuclear potential in hot and dense matter}

\author{Laura Tol\'os, Angels Ramos, Artur Polls}
\address{Departament d'Estructura i Constituents de la Mat\`eria,
Universitat de Barcelona, \\
Diagonal 647, 08028 Barcelona, Spain
}

\date{\today}

\maketitle

\begin{abstract}
The antikaon optical potential in hot and dense nuclear matter is
studied within the framework of a coupled-channel self-consistent
calculation taking, as bare meson-baryon interaction, the
meson-exchange potential of the J\"ulich group. Typical conditions
found in heavy-ion collisions at GSI are explored. As in the case
of zero temperature, the angular momentum components larger than
$L=0$ contribute significantly to the finite temperature antikaon
optical potential at finite momentum. It is found that the
particular treatment of the medium effects has a strong influence
on the behavior of the antikaon potential with temperature. Our
self-consistent model, in which antikaons and pions are dressed in
the medium, gives a moderately temperature dependent antikaon
potential which remains attractive at GSI temperatures, contrary
to what one finds if only nuclear Pauli blocking effects are
included.

\vspace{0.5cm}



\end{abstract}

%\maketitle

\section{Introduction}
\label{sec:intro}

The study of the properties of hadrons in hot and dense matter is
receiving a lot of attention in the last years, in connection to
the experimental programs at SIS-GSI, SPS-CERN, RHIC-BNL and the
forthcoming operation of LHC \cite{experiments}. A particular
effort has been invested in understanding the properties of
antikaons due to the direct implications they have in
astrophysical phenomena. If the antikaon mass drops in matter
below the electron chemical potential, a condensed fraction of
$K^-$ would appear \cite{KN86}, giving rise to a softer nuclear
equation of state and to a substantial reduction of the maximum
mass that neutron stars could sustain (see updated references in
Ref.~\cite{kaon}).

There have been some attempts to extract the antikaon-nucleus
potential from best-fit analysis of kaonic-atom data and some
solutions favor very strongly attractive well depths of the order
of 200 MeV at normal nuclear matter density \cite{FGB94}. However,
recent self-consistent calculations based on a chiral lagrangian
\cite{Lutz98,Oset00,Schaffner} or meson-exchange potentials
\cite{Laura} only predict moderate attractive depths of 50 to 80
MeV. In addition, studies of kaonic atoms using the chiral ${\bar
K}N$ amplitudes of Ref.~\cite{oset98} show that it is indeed
possible to find a reasonable reproduction of the data with a
relatively shallow antikaon-nucleus potential \cite{satoru00},
albeit adding an additional moderate phenomenological piece
\cite{baca}. This has been recently corroborated by a calculation
\cite{cieply01}, where a good fit to both scattering $K^- p$ data
and kaonic-atom data only required to modify slightly the
parameters of the chiral meson-baryon interaction model of
Ref.~\cite{weise}. The lesson learned is that kaonic atom data do
not really provide a suitable constraint on the antikaon-nucleus
potential at normal nuclear matter density.

Heavy-ion collisions can also offer information on the modification
of the properties of hadrons. In particular, the enhanced $K^-/K^+$ ratio
measured at GSI \cite{kaos97} can  be understood assuming that the
antikaons feel a strong attraction \cite{cassing97,Li97,Li98,sibirt98},
although other interpretations in terms of the
enhanced cross sections due to the
shifting of the $\Lambda(1405)$ resonance to higher energies have also
been advocated \cite{Schaffner}, a mechanism
that was already suggested in Ref.~\cite{Ohnishi97} to explain
the data of $K^-$ absorption at rest on $^{12}$C
\cite{Tamura89,Kubota96}.

In fact, the antikaon-nucleon amplitude in isospin $I=0$ is
dominated by the presence of this $\Lambda(1405)$ resonant
structure, which in free space appears only 27 MeV below the
${\bar K}N$ threshold. The resonance is generated dynamically from
a $T$-matrix scattering equation in coupled channels using a
suitable meson-baryon potential. The coupling between the ${\bar
K}N$ and $\pi Y$ ($Y=\Lambda,\Sigma$) channels is essential to get
the right dynamical behavior in free space. Relatedly, the
in-medium properties of the $\Lambda(1405)$ such as its position
and its width, which in turn influence strongly the behavior of
the antikaon-nucleus optical potential, are very sensitive to the
many-body treatment of the medium effects. Previous works have
shown, for instance, that a self-consistent treatment of the
$\bar{K}$ self-energy had a strong influence on the scattering
amplitudes \cite{Lutz98,Oset00,Schaffner} and, consequently, on
the in-medium properties of the antikaon. The incorporation of the
pion with its medium modified properties also proved to be
important \cite{Oset00}, although most works until now have
ignored it \cite{Lutz98,Schaffner,Laura,cieply01}.

In our previous work \cite{Laura} we performed a study of the
antikaon properties in the medium within a self-consistent
dynamical model which used, as bare interaction, the
meson-exchange potential of the J\"ulich group \cite{holinde90}.
Since antikaons are produced at finite momentum, we focussed on
the effect of high partial waves, typically excluded in the chiral
models \cite{Lutz98,Oset00}. We note, however, that 
the p-waves
have been incorporated recently in the chiral amplitudes
\cite{caro,Lutz01}, and a first estimation of their contribution
to the in-medium properties of the antikaons has also been recently reported \cite{Lutz011}.

One must keep in mind that heavy-ion collisions produce dense and
hot matter. Therefore, to learn about the meson properties in this
environment, it is necessary to incorporate in the formalism not
only the medium effects but also those associated with having a
finite non-zero temperature. The purpose of the present work is,
precisely, to develop a proper self-consistent calculation of the
antikaon properties in a dense and hot medium, extending to finite
$T$ our previous $T=0$ model \cite{Laura}. We will indeed show
that the antikaon properties at finite $T$ depend very strongly on
the treatment of the medium. The present work focuses on typical
scenarios found in heavy-ion collisions at GSI. For this reason,
results will be shown up to densities of twice nuclear saturation
density and up to temperatures of 70 MeV.

We have organized the present paper as follows. In
Sect.~\ref{sec:formalism} we first review our formalism at $T=0$
and include some updated ingredients, such as the incorporation of
the pion self-energy and the use of some slightly different
prescriptions for the single-particle energies of the baryons.
Next, we show how we treat the temperature effects, which amount
to using the appropriate finite $T$ nucleon Pauli blocking factor,
as well as using temperature dependent self-energies for the
mesons (antikaons and pions) and the nucleons. Some lengthy
derivations are relegated to the appendix. Our results are
presented and discussed in Sect.~\ref{sec:results} and the
concluding remarks are given in Sect.~\ref{sec:conclusions}.

\section{FORMALISM}
\label{sec:formalism}

In this section we review the calculation of the single-particle
potential of the $\bar{K}$ meson at $T=0$ in symmetric nuclear
matter.  In addition to the self-consistent incorporation of the
$\bar{K}$ self-energy, as done in our previous work \cite{Laura},
we also include here the self-energy of the pion in the
intermediate $\pi\Lambda$, $\pi\Sigma$ states present in the
construction of the $\bar{K}N$ effective interaction, since this
was shown to be important in Ref.~\cite{Oset00}. In a further
step, we extend the model to finite temperature, which affects the
Pauli blocking of the nucleons in the intermediate states, as well
as the dressing of mesons and baryons.

\subsection{Revision of the $T=0$ formalism}

In Ref.~\cite{Laura}, the effective $\bar{K} N$ interaction in the
nuclear medium ($G$-matrix) at $T=0$ was derived from a
meson-baryon bare interaction built in the meson exchange
framework \cite{holinde90}. As the bare interaction permits
transitions from the $\bar {K} N$ channel to the $\pi \Sigma$ and
$\pi \Lambda$ ones, all having strangeness $S=-1$, we are
confronted with a coupled channel problem.  Working in isospin
coupled basis, the $\bar {K} N$ channel can have isospin $I=0$ or
$I=1$, so the resultant $G$-matrices are classified according to
the value of isospin. For $I=0$, $\bar {K} N$ and $\pi \Sigma$ are
the only channels available, while for $I=1$ the $\pi \Lambda$
channel is also allowed. In a schematic notation, each $G$-matrix
fulfills the coupled channel equation

\begin{eqnarray}
\langle M_1 B_1 \mid G(\Omega) \mid M_2 B_2 \rangle &&=
\langle M_1 B_1
\mid V({\sqrt s}) \mid M_2 B_2 \rangle   \nonumber \\
&& \hspace*{-2cm}+\sum_{M_3 B_3} \langle M_1 B_1 \mid V({\sqrt
s }) \mid
M_3 B_3 \rangle
\frac {Q_{M_3 B_3}}{\Omega-E_{M_3} -E_{B_3}+i\eta} \langle M_3
B_3 \mid
G(\Omega)
\mid M_2 B_2 \rangle \ ,
   \label{eq:gmat1}
\end{eqnarray}
where $\Omega$ is the so-called starting energy, given in the lab
frame, and $\sqrt{s}$ is the invariant center-of-mass energy. In
Eq.~(\ref{eq:gmat1}),  $M_i$ and $B_i$  represent, respectively,
the possible mesons ($\bar {K}$, $\pi$) and baryons ($N$,
$\Lambda$, $\Sigma$), and their corresponding quantum numbers,
such as coupled spin and isospin, and linear momentum. The
function $Q_{M_3 B_3}$ stands for the Pauli operator preventing
the nucleons in the intermediate states from occupying already
filled ones.

The  prescription for the
single-particle energies of all the mesons and baryons
participating
in the reaction and in the intermediate states is written as

\begin{equation}
 E_{M_i(B_i)}(k)=\sqrt{k^2 +m_{M_i(B_i)}^2} + U_{M_i(B_i)}
(k,E_{M_i(B_i)}^{qp}) \ ,
\label{eq:spen}
\end{equation}
where $U_{M_i(B_i)}$ is the  single-particle
potential of each meson (baryon) calculated at the real
quasi-particle energy $E_{M_i(B_i)}^{qp}$. For baryons, this
quasi-particle energy is given by

\begin{equation}
E_{B_i}^{qp}(k)=\sqrt{k^2 +m_{B_i}^2} +  U_{B_i} (k) \ ,
\label{eq:qpb}
\end{equation}
while, for mesons, it is obtained by solving the following equation
\begin{equation}
(E_{M_i}^{qp}(k))^{2}=k^2 +m_{M_i}^2 + {\mathrm {Re}\,} \Pi_{M_i}
(k,E_{M_i}^{qp}) \ ,\label{eq:qpm}
\end{equation}
where $\Pi_{M_i}$ is the meson self-energy.

The $\bar K$ single-particle potential in the Brueckner-Hartree-Fock
approach is schematically given by
\begin{equation}
 U_{\bar K}(k_{\bar{K}},E_{\bar K}^{qp})= \sum_{N \leq F} \langle \bar K
N \mid
 G_{\bar K N\rightarrow
\bar K N} (\Omega = E^{qp}_N+E^{qp}_{\bar K}) \mid \bar K N
\rangle \ ,
\label{eq:self0}
\end{equation}
where the summation over nucleon states is limited by the nucleon
Fermi momentum. The ${\bar K}$ self-energy is obtained from the
optical potential through the relation
\begin{equation}
\Pi_{\bar{K}}(k_{\bar{K}},\omega)=2 \
\sqrt{k_{\bar{K}}^2+m_{\bar{K}}^2} \
U_{\bar{k}}(k_{\bar{K}},\omega) \ .
\label{eq:pik}
\end{equation}
As it can be easily seen from Eq.~(\ref{eq:self0}), since the
$\bar{K}N$ effective interaction ($G$-matrix) depends on the
$\bar{K}$ single-particle energy, which in turn depends on the
${\bar K}$ potential through Eq.~(\ref{eq:pik}), we are confronted
with a self-consistency problem.

We proceed as in Ref.~\cite{Laura}, where self-consistency for the
optical potential was demanded at the quasi-particle energy. After
self-consistency is reached, the complete $\bar{K}$
energy-and-momentum dependent self-energy of the ${\bar K}$ can be
obtained, from which the $\bar K$ propagator
\begin{equation}
D_{\bar{K}}(k_{\bar{K}},\omega)=
\frac{1}{\omega^2-k_{\bar{K}}^2-m_{\bar{K}}^2-
\Pi_{\bar{K}}(k_{\bar{K}},\omega)} \ , \label{eq:prop}
\end{equation}
and the corresponding spectral density, defined as
\begin{equation}
S_{\bar K}(k_{\bar K},\omega) = - \frac {1}{\pi} {\mathrm Im\,} D_{\bar
K}(k_{\bar K},\omega) \ ,
\label{eq:spec}
\end{equation}
can be derived. We note that our self-consistent procedure amounts
to replacing in the ${\bar K}$ propagator the energy dependent
self-energy, $\Pi_{\bar{K}}(k_{\bar{K}},\omega)$, by that
evaluated at the quasiparticle energy,
$\Pi_{\bar{K}}(k_{\bar{K}},\omega=E^{qp}_{\bar K}(k_{\bar K}))$.
This is what we refer to as the quasi-particle self-consistent
approach, which retains the position and the width of the peak of
the ${\bar K}$ spectral function at each iteration. This
simplification versus the more complete self-consistent job done
in Refs.~\cite{Lutz98,Oset00} allows one to perform analytically
the energy integral of the intermediate loops, thus reducing the
four-dimensional integral equation to a three-dimensional one.

In this paper, our model at $T=0$ has been reviewed by implementing
some modifications in the properties of the baryons and mesons.
Firstly, we have introduced slight modifications in the
single-particle potential of the $\Lambda$ and $\Sigma$ hyperons,
following the parameterization of Ref.~\cite{Gal97},
\begin{equation}
U_{\Lambda, \Sigma} (\rho) = -340 \rho + 1087.5 \rho^2 \ ,
\end{equation}
which is more appropriate in the high density regime than the
simple parameterization, linear in $\rho$, used previously.

For nucleons, we have used a relativistic $\sigma-\omega$ model,
where the scalar and vector coupling constants, $g_s$ and $g_v$
respectively, are density dependent \cite{Mach89}. This is a
simple way to mimic results from Dirac-Brueckner-Hartree-Fock
calculations.

The most important modification comes from the introduction of the
pion self-energy, $\Pi_{\pi}(k_{\pi},\omega)$, in the intermediate
$\pi\Sigma$, $\pi\Lambda$ states present in the construction of
the $\bar{K}N$ effective interaction. The pion is dressed with the
momentum and energy-dependent self-energy already used in
Ref.~\cite{Oset00}. This self-energy incorporates a p-wave piece
built up from the coupling to $1p-1h$, $1\Delta-1h$ and $2p-2h$
excitations, plus short-range correlations. The model also
contains a small and constant s-wave part (see the appendix for
details and appropriate references of the model).

As in the case of the $\bar{K}$ meson, we can define a pion
optical potential, $U_{\pi}(k_{\pi},\omega)$, from the complete
$\pi$ self-energy. We use
\begin{equation}
U_{\pi}(k_{\pi},\omega)=\sqrt{k_{\pi}^2+m_{\pi}^2+
\Pi_{\pi}(k_{\pi},\omega)}-\sqrt{k_{\pi}^2+m_{\pi}^2} \ ,
\label{eq:upi}
\end{equation}
which ensures that, when inserted in the $G$-matrix equation, one
is using a quasiparticle approximation to the spectral function,
with the peak located at the right quasiparticle energy defined in
Eq.~(\ref{eq:qpm}). Due to the small mass of the pion, it has not
been possible to retain only the first term in the expansion of
the first square root of Eq.~(\ref{eq:upi}), as we did for the
$\bar{K}$ meson.

The spectral density for $\pi$ is calculated from
Eqs.~(\ref{eq:prop}) and (\ref{eq:spec}), by replacing the
$\bar{K}$ properties for those of the $\pi$.

\subsection{Temperature effects}

The introduction of temperature in the calculation of the
$G$-matrix affects the Pauli blocking of the intermediate nucleon
states, as well as the dressing of mesons and baryons. The
$G$-matrix equation at finite $T$ reads formally as in
Eq.~(\ref{eq:gmat1}), but replacing

\begin{eqnarray}
Q_{M B} &\rightarrow& Q_{M B}(T) \nonumber \\
G(\Omega) &\rightarrow& G(\Omega,T) \nonumber \\
E_M \ , E_B &\rightarrow& E_M(T) \ , E_B(T) \nonumber \ .
\end{eqnarray}
The function $Q_{M B}(T)$ is unity for meson-hyperon states while,
for $\bar{K}N$ states, it follows the law $1-n(k_N,T)$, where
$n(k_N,T)$ is the nucleon Fermi distribution at the corresponding
temperature
\begin{eqnarray}
n(k_N,T)=\frac{1}{1+{\rm
e}^{\left(\frac{E_N(k_N,T)-\mu}{T}\right)}} \ .
\label{eq:fermidistribution}
\end{eqnarray}
The nucleonic spectrum at finite $T$,
$E_N(k_N,T)$, is obtained following a
$\sigma-\omega$ model that will be described below and $\mu$ is the
chemical potential obtained imposing the normalization property
\begin{eqnarray}
\rho=\frac{\nu}{(2\pi)^3} \int  d^3k_N  \ n(k_N,T) \ ,
\label{eq:density}
\end{eqnarray}
at each density $\rho$, where $\nu=4$ is the degeneration factor of
symmetric nuclear matter.

As in the $T=0$ case, we perform an angle average
of the Pauli operator $Q_{\bar{K}N}(T)$, a strategy which facilitates the
solution of the $G$-matrix equation in a partial wave basis. We first
define $\vec{P}$ and $\vec{k}$ as the total and relative momenta of
the $\bar{K}N$ pair, respectively,
\begin{equation}
\vec{P}=\vec{k}_{\bar{K}}+\vec{k}_{N} ,~~~~
\vec{k}=\frac{m_{N}\vec{k}_{\bar{K}}-m_{\bar{K}}\vec{k}_{N}}
{m_{\bar{K}}+m_{N}} \ ,
\end{equation}
which allow us to rewrite the nucleon and antikaon momenta in the
laboratory system, $\vec{k}_{N}$ and $\vec{k}_{\bar{K}}$, as
\begin{equation}
\vec{k}_{N}=-\vec{k}+\frac{\xi}{1+\xi}\vec{P} , ~~~~
\vec{k}_{\bar{K}}=\vec{k}+\frac{1}{1+\xi}\vec{P} \ ,
\end{equation}
where $m_{\bar K}$, $m_N$ are the kaon, nucleon masses,
respectively, and $\xi=\displaystyle\frac{m_{N}}{m_{\bar{K}}}$. In
terms of the total and relative momenta, the Pauli operator
$Q_{{\bar K}N}(k_N,T)$ reads $Q_{{\bar K}N}
(\mid\frac{\xi}{1+\xi}\vec{P}-\vec{k}\mid,T)$  , which shows
explicitly the dependence on the angle between $\vec{P}$ and
$\vec{k}$. This dependence is eliminated in the $G$-matrix
equation by replacing $Q_{{\bar K}N}$ by its angle average
\begin{equation}
\overline{Q}_{{\bar K}N}(P,k,T)=\frac{1}{2} \int_{-1}^{1} dx \
Q_{{\bar K}N}(\vec{P},\vec{k},T)= \frac{1}{2B} \ \ln \frac{ {\rm
e}^A+{\rm e}^B}{{\rm e}^A+{\rm e}^{-B}} \ ,
\end{equation}
where
\begin{eqnarray}
A&=&\frac{\mu}{T}-\frac{k^2+({\frac{\xi}{\xi+1}})^2 P^2}{2 m_N T}
\nonumber \\
B&=&\frac{k \ \xi \ P }{T \ (1+\xi) \ m_N} \ .
\end{eqnarray}

Temperature also affects the properties of the particles involved in the process.
The $\bar K$ optical potential at a given temperature  is
calculated according to

\begin{equation}
 U_{\bar K}(k_{\bar{K}},E_{\bar K}^{qp},T)= \int d^3k_N \ n(k_N,T) \ \langle \bar K N \mid
 G_{\bar K N\rightarrow
\bar K N} (\Omega = E^{qp}_N+E^{qp}_{\bar K},T) \mid \bar K N \rangle \ .
\label{eq:self}
\end{equation}
Once more, this is a self-consistent problem for $U_{\bar{K}}$.
More explicitly, using the partial wave components of the
$G$-matrix, we obtain (similarly to Ref.~\cite{Laura})
\begin{eqnarray}
 U_{\bar{K}}&&(k_{\bar{K}},E_{\bar K}^{qp},T)=
 \frac{1}{2}
\sum_{L, J, I}(2J+1)(2I+1) \int n(k_N,T)\  k_N^2 \ dk_N \\ \nonumber
&& \times  \langle (\bar{K}N) ; \overline{k}| G^{L J I}
(\overline{P},E^{qp}_{\bar{K}}(k_{\bar{K}})+
E^{qp}_{N}(k_{N}),T)
|
(\bar{K} N); \overline{k} \rangle  \ ,
\label{eq:upot1}
\end{eqnarray}
where $\overline{k}$ and $\overline{P}$ are the  relative and
center-of-mass momentum, respectively, averaged over the angle
between the external $\bar{K}$ momentum in the lab system,
$k_{\bar{K}}$, and the internal momentum of the nucleon, $k_N$. In the actual calculations, we include partial waves up to $L=4$.

We also have to pay attention to the temperature effects on the
properties of the other hadrons participating in the process. We
have to be especially careful with the pion, since its small  mass
makes it very sensitive to variations in its properties. As
mentioned before, the pion  self-energy at $T=0$ has been obtained
following a model that includes the coupling to $1p-1h$,
$1\Delta-1h$ and $2p-2h$ excitations. The details on how this
model is modified by the effect of a finite temperature are
presented in the appendix.

In the case of nucleons, we have introduced temperature effects
following the Walecka $\sigma-\omega$ model \cite{Adv}, using the
density dependent scalar and vector coupling constants at $T=0$
given in Ref.~\cite {Mach89}. We have obtained the baryonic
chemical potential, $\mu$, the effective mass, $m^*$, and the
vector potential, $\Sigma^0$, simultaneously, fixing the
temperature and the density. The value of $\mu$ is defined via the
baryonic density, $\rho$, [see Eqs.~(\ref{eq:fermidistribution})
and (\ref{eq:density})], where $E_N(k_N,T)$ is the nucleonic
energy spectrum,
\begin{eqnarray}
E_N(k_N,T)=\sqrt{k_N^2+m^*(T)^2}-\Sigma^0 \ ,
\end{eqnarray}
and $\Sigma_0$ and $m^*$ are defined as
\begin{eqnarray}
\Sigma^0&=&-\left(\frac{g_v}{m_v}\right)^2 \rho    \nonumber \\
m^*(T)&=&m-\left(\frac{g_s}{m_s}\right)^2\frac{\nu}{(2\pi)^3} \int
d^3k_N \ \frac{m^*(T)}{\sqrt{k_N^2+m^*(T)^2}} \  n(k_N,T) \ .
\end{eqnarray}
One clearly sees that,
given $g_s$ and $g_v$ at $T=0$, and a fixed $\rho$, $\Sigma^0$ is easily
obtained. On the other hand,
a simultaneous solution of $\mu$ and $m^*$ is needed
to determine the nucleonic spectrum.

We note that the antiparticle contribution to the density $\rho$
has been proven to be negligible, so it is not considered here. We
have not considered either changes in the hyperon properties
induced by the use of a finite temperature. We have checked, in a
schematic Skyrme-Hartree-Fock calculation, that these changes are
small within the temperature range explored in this work.

\section{Results}
\label{sec:results}

\begin{figure}[htb]
\centerline{
     \includegraphics[width=0.6\textwidth]{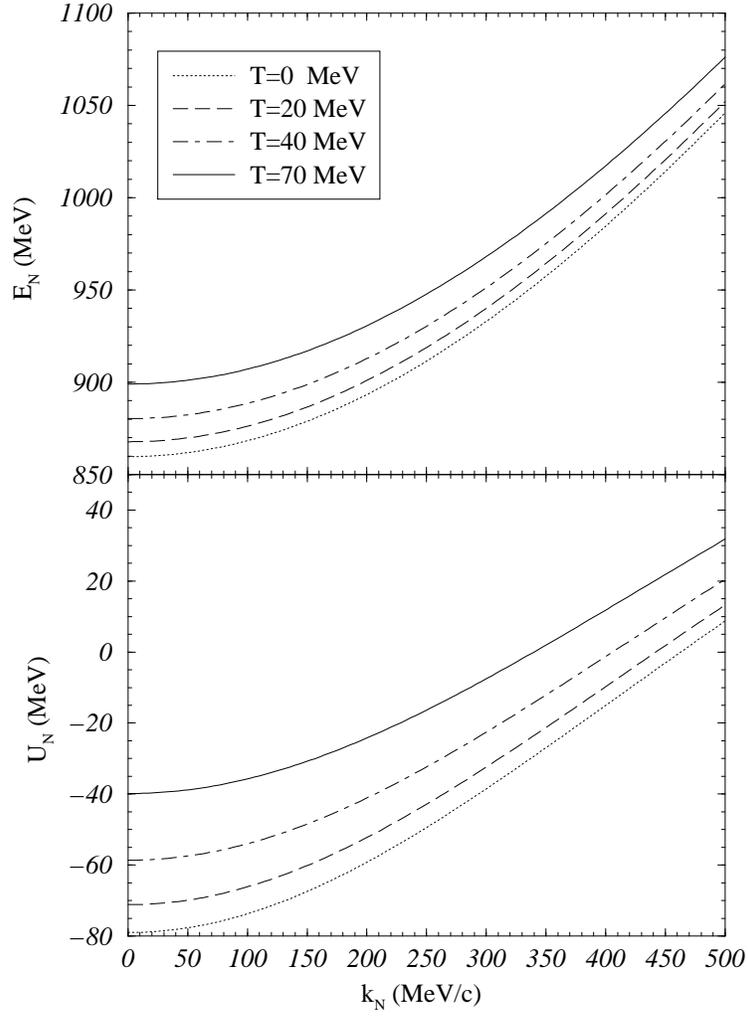}
}
      \caption{\small Nucleon energy spectrum and nucleon
      potential at $\rho=0.17 \ \rm{fm^{-3}}$ as function of the nucleon momentum
for different temperatures. }
        \label{fig:temp1}
\end{figure}

We start this section by studying the effect of the temperature on
the nucleon spectrum and the pion self-energy, both of them
crucial ingredients for the calculation of $U_{\bar{K}}$. In
Fig.~\ref{fig:temp1}, the nucleon spectrum and the nucleon
potential, defined as $U_N=E_N-\sqrt{k_N^2+m_N^2}$, are given as
functions of the nucleon momentum for various temperatures at the
nuclear saturation density $\rho_0=0.17 \ \rm{fm}^{-3}$. The
lowest curves in both graphs correspond to $T=0$ and´, as the
value of the temperature increases up to $T=70$ MeV (solid lines),
the attractive potential gets reduced from around $-80 \ \rm{MeV}$
to $-40 \ \rm{MeV}$ at $k_N=0$. Consequently, the energy spectrum
for $k_N=0$ goes from $860 \ \rm{MeV}$ to $900 \ \rm{MeV}$. A
similar trend is observed for higher values of the nucleon
momentum. This behavior is well known for the $\sigma-\omega$
models, as it is reported in Ref.~\cite{Adv}.

\begin{figure}[htb]
\centerline{
     \includegraphics[width=0.6\textwidth,angle=-90]{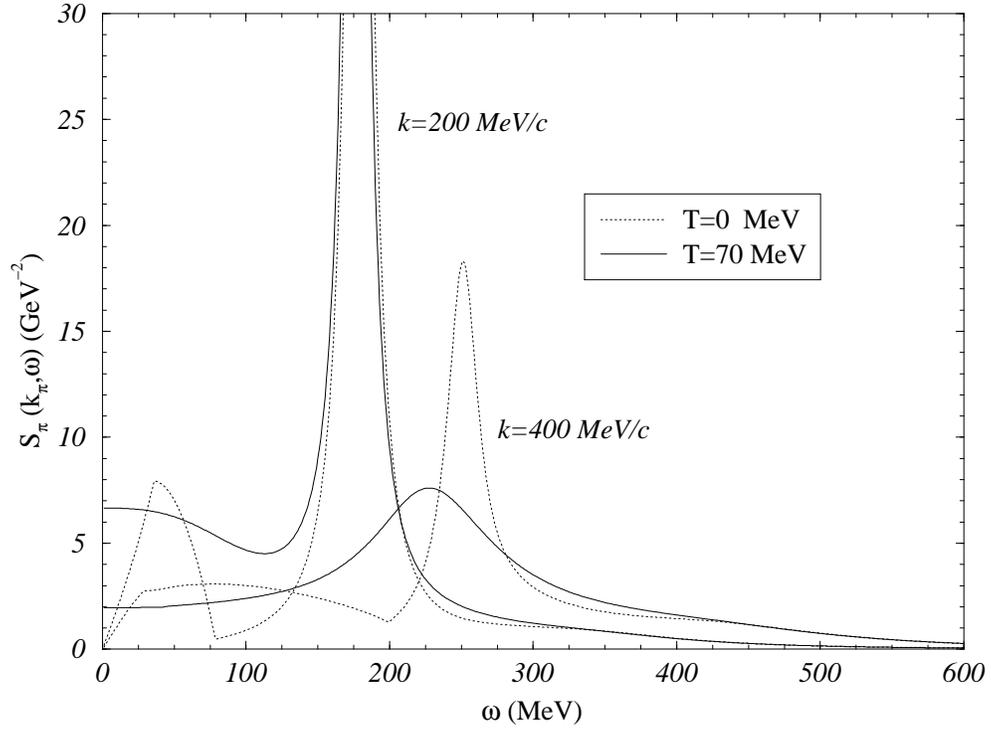}
}
      \caption{\small Spectral density of the pion at
      $\rho=0.17 \ \rm{fm^{-3}}$ as a function of
energy for $k_{\pi}=200 \ \rm{MeV}/c$ and $k_{\pi}=400 \
\rm{MeV}/c$, and for $T=0 \ \rm{MeV}$ and $T=70 \ \rm{MeV}$. }
        \label{fig:temp2}
\end{figure}

The next two figures give us information about the pion
self-energy and how it is modified by temperature. Figure
\ref{fig:temp2} displays the spectral density of the pion as a
function of energy at nuclear saturation density, $\rho_0=0.17 \
\rm{fm}^{-3}$, for two pion momenta, $k_{\pi}=200, 400 \
\rm{MeV}$, and two different temperatures, $T=0 \ \rm{MeV}$
(dotted lines) and $T=70 \ \rm{MeV}$ (solid lines). At $T=0$, the
structure of the $1p-1h$ excitations can be seen very clearly on
the left side of the quasiparticle peak. This structure smooths
out as temperature increases. The effect of $1\Delta-1h$
excitations is more difficult to be identified for these two
momenta. It is only somewhat appreciated at $T=0$ by a slower fall
of the spectral function with energy to the right of the
quasiparticle peak. The effect of temperature is to move the
quasiparticle peak slightly away from the $T=0$ position towards
lower energies, making it noticeably wider.

\begin{figure}[htb]
\centerline{
     \includegraphics[width=0.6\textwidth]{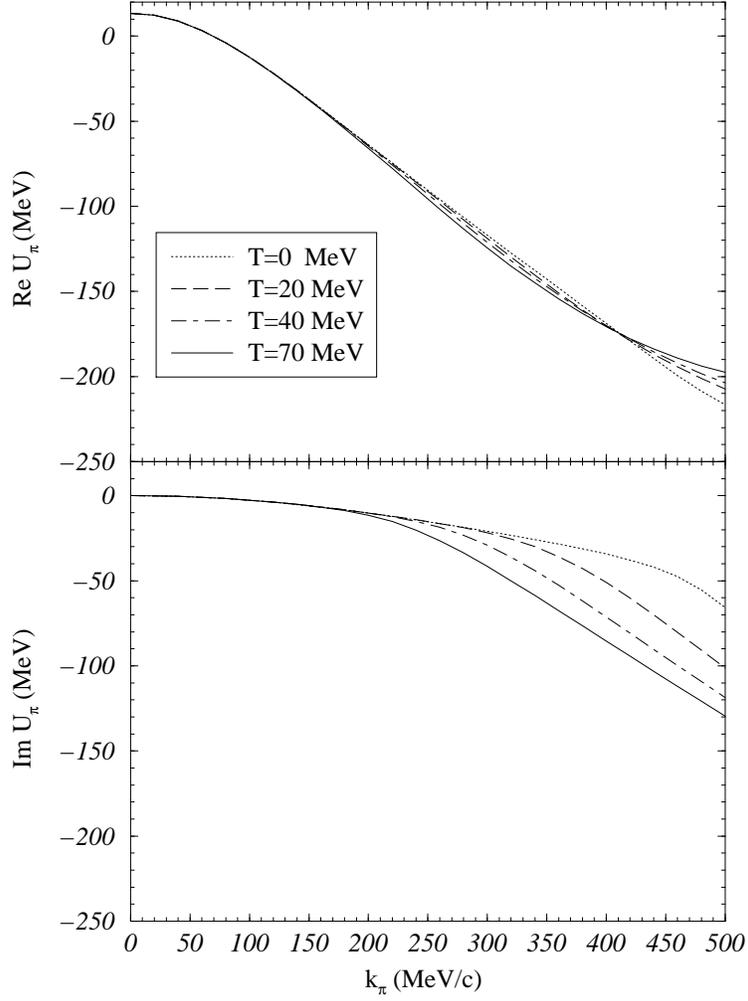}
}
      \caption{\small
Real and imaginary parts of $U_{\pi}$ at $\rho=0.17 \
\rm{fm^{-3}}$  as functions of the pion momentum for various
temperatures.}
        \label{fig:temp3}
\end{figure}

In Fig.~\ref{fig:temp3} we show the pion optical potential as
defined in Eq.~(\ref{eq:upi}). The real and imaginary parts of the
pion optical potential at nuclear saturation density are displayed
as functions of the pion momentum, $k_{\pi}$, for different
temperatures. The dotted lines correspond to $T=0$, and the
results for the highest temperature studied, $T=70 \ \rm{MeV}$,
are represented by the solid ones. In the region of pion momenta
explored here, the imaginary part shows a stronger dependence on
the temperature than the real part, which is practically
$T$-independent. Note that the quasiparticle energy, which defines
the position of the quasiparticle peak in the spectral function,
is determined through the real part of the pion self-energy [see
Eq.~(\ref{eq:qpm})], and it is not directly obtained from
${\rm Re}\, U_{\pi}$, where $U_{\pi}$ is given in
Eq.~(\ref{eq:upi}). This explains that, while ${\rm Re}\, U_{\pi}$
at $k_{\pi}=400$ MeV is practically the same for $T=0$ and $70$
MeV, the location of the peak differs more markedly. Similarly,
the width (or height) of the peak for the different temperatures
cannot be directly calculated from ${\rm Im}\, U_{\pi}$, but has to
be obtained from ${\rm Im}\, \Pi_{\pi}$ at the quasiparticle
energy.

\begin{figure}[htb]
\centerline{
     \includegraphics[width=0.6\textwidth]{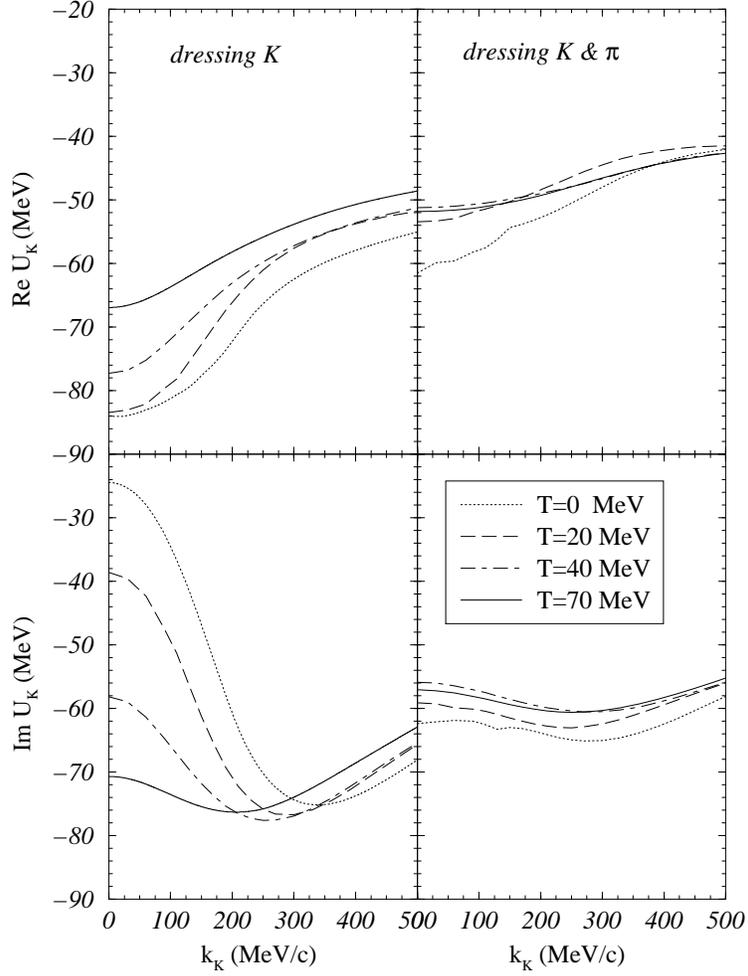}
}
      \caption{\small Real and imaginary parts of
      $U_{\bar{K}}$ at $\rho=0.17 \ \rm{fm^{-3}}$
      as functions of the antikaon momentum for
several temperatures, dressing only $\bar{K}$ (left panels) and
dressing both $\bar{K}$ and $\pi$ (right panels). }
        \label{fig:temp4}
\end{figure}

\begin{figure}[htb]
\centerline{
     \includegraphics[width=0.6\textwidth,angle=-90]{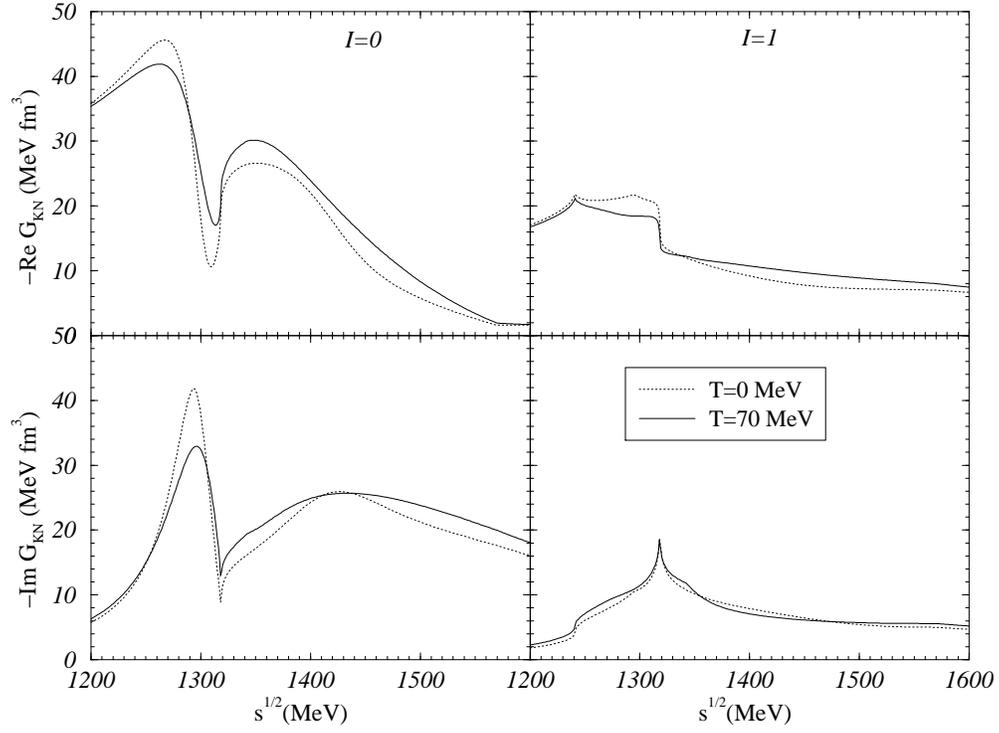}
}
      \caption{\small
Real and imaginary parts of the ${\bar K}N$ amplitude in the
$I=0$, $L=0$ channel (left panels) and the $I=1$, $L=0$ channel
(right panels) at $\rho=0.17 \ \rm{fm^{-3}}$ as functions of the
center-of-mass energy at total momentum $|\vec k_{\bar K} +\vec
k_{N}|=0$ for $T=0 \ \rm{MeV}$ and $T=70 \ \rm{MeV}$.}
        \label{fig:temp5}
\end{figure}

Once the pion self-energy is introduced in the calculation of the
$\bar{K}$ optical potential, its effects can be studied by
comparing the results obtained by dressing only the $\bar{K}$
mesons with those in which not only the $\bar{K}$ mesons but also
the pions in the intermediate meson-baryon states are dressed. In
Fig. \ref{fig:temp4}, the real and imaginary parts of the
$\bar{K}$ optical potential at nuclear saturation density are
shown as functions of the antikaon momentum, $k_{\bar{K}}$, for
different temperatures. On the left panels, only the $\bar{K}$
mesons have been dressed, while the results on the right panels
incorporate, in addition, the dressing of the pions. Dotted lines
correspond to $T=0$ and solid lines to $T=70 \ \rm{MeV}$. At $T=0$
we find the same qualitative effects from dressing the pions as
those found by the chiral model \cite{Oset00} shown in
Ref.~\cite{procTorino}. When the pions are dressed, the real part
of the antikaon potential becomes less attractive and the
imaginary part loses structure. This behavior is a direct
consequence of a smoother energy dependence of the ${\bar K}N$
effective interaction when the pions are dressed [see Fig.~4 in
Ref.~\cite{Oset00}], as well as to the fact that, due to the less
attractive antikaon potential, one explores this interaction at
higher energies, further away from the resonant structure. When
temperature increases, the optical potential loses attraction and
absorptive power, although the effect is moderate. The reason is
obvious: as it can be seen from Eq.~(\ref{eq:self}), and assuming
a weak dependence of the effective interaction $G_{\bar{K}N}$ on
temperature, at a non-vanishing $T$, one is averaging over higher
momentum states, where this interaction is weaker. Nevertheless,
at sufficient high $T$ [see the $70 \ \rm{MeV}$ results on the
right panels], one starts to gain some attraction and absorption.
Evidently, this is a consequence of the $T$-dependence of the
effective interaction. To visualize this additional dependence on
$T$, we show in Fig. \ref{fig:temp5}, the effective ${\bar K}N$
interaction for the channels $L=0$, $I=0$ and $L=0$, $I=1$, at
zero center-of-mass momentum, for two different temperatures,
$T=0$ (dotted lines) and $T=70 \ \rm{MeV}$ (solid lines). The
general trends of the effective ${\bar K}N$ interaction at these
two temperatures are similar but there are small differences that
explain the behavior observed in the previous figure for the
${\bar K}$ optical potential. Indeed, the ${\bar K}N$ amplitudes
involved in the construction of the ${\bar K}$ optical potential
correspond to energies above 1300 MeV and, in that region, the
magnitude of the real part of the $T=70$ amplitudes is larger than
the $T=0$ ones, thereby compensating the loss of attraction
induced by the higher relative momentum components present in the
${\bar K}$ optical potential at finite $T$.

It is especially interesting to observe the structure in the $I=0$
amplitude appearing below the $\pi\Sigma$ threshold, a region in
energy not explored in previous works \cite{Lutz98,Oset00,Laura}.
It appears that the resonance in the medium, previously identified
with the bump in the imaginary part around 1400 MeV, might be more
appropriately identified with the more pronounced peak appearing
around 1300 MeV. Whether this is a new resonance (an additional
pole in the complex plane) or just a reflection of the same one
but distorted by the presence of a cusp at the $\pi\Sigma$
threshold is something that deserves further study. At the moment,
we can only say that there is an enhanced probability of finding a
state with $\Lambda$-like quantum numbers around 1300 MeV. We note
that this would have been signaled by a pole in the real axis if
neither the pions nor the antikaons would have been dressed, since
in this case there would not have been allowed states to decay to.
In the self-consistent many-body approach used here, the states to
which this peaked structure can decay are of the type
$\pi(ph)\Sigma$ or ${\bar K}(\Lambda h \pi) N$, where in
parentheses we have denoted an example for the component of the
$\pi$ and ${\bar K}$ mesons that could show up at energies below
the $\pi \Sigma$ threshold.

\begin{figure}[htb]
\centerline{
     \includegraphics[width=0.6\textwidth]{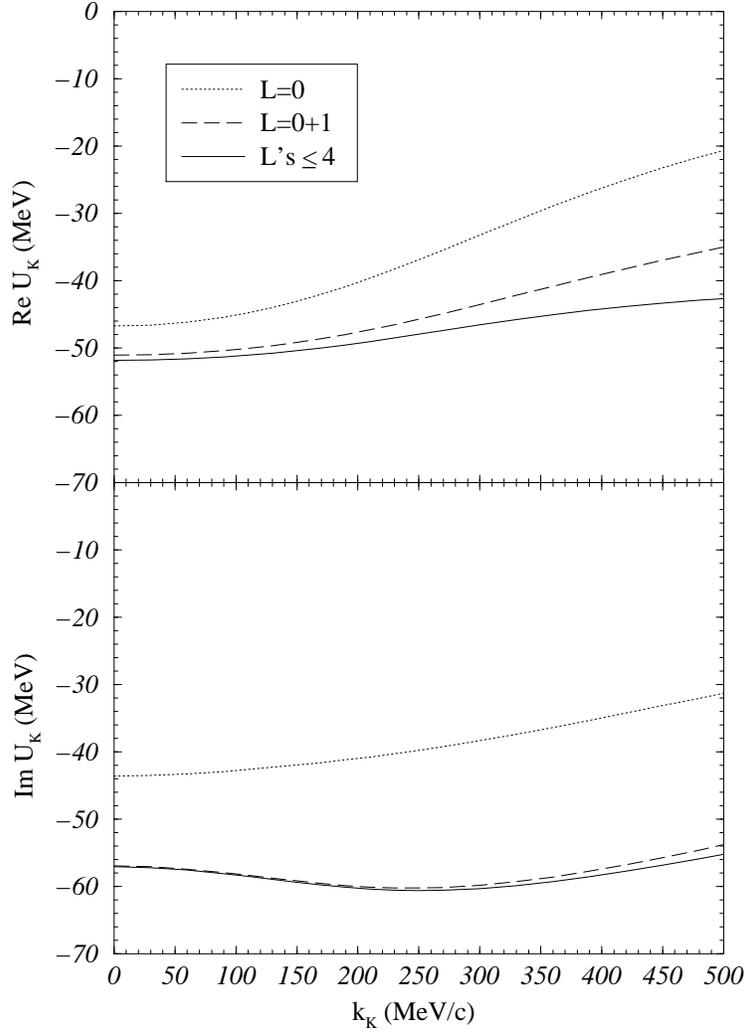}
}
      \caption{\small
      Partial wave contributions to the
real and imaginary parts of the ${\bar K}$ optical potential at
$\rho=0.17 \ \rm{fm^{-3}}$ as functions of the antikaon momentum
for $T=70 \ \rm{MeV}$.}
        \label{fig:temp6}
\end{figure}

We have also studied the contributions of angular momentum
components larger than $L=0$ to the antikaon optical potential, as
we did in our $T=0$ work \cite{Laura}. In Fig. \ref{fig:temp6}, we
display the contribution of the different partial waves to the
real and imaginary parts of the $\bar{K}$ optical potential at
nuclear saturation density and a temperature of $T=70 \ \rm{MeV}$.
We observe that, adding the higher partial waves to the $L=0$
contribution, produces significant changes. The momentum dependence
of the $\bar{K}$ optical potential becomes smoother, the real part
becomes more attractive, and the imaginary part increases by about
$25\%$ at $k_{\bar{K}}=0$ and by $50\%$ at momenta around $500 \
\rm{MeV/c}$.

\begin{figure}[htb]
\centerline{
     \includegraphics[width=0.6\textwidth]{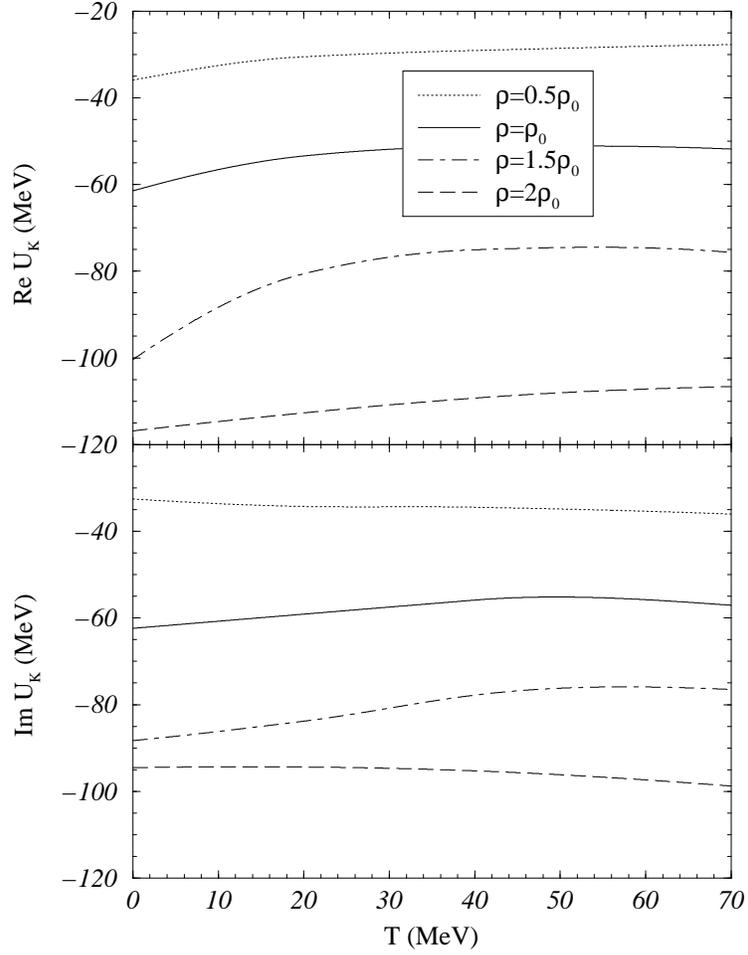}
}
      \caption{\small
Real and imaginary parts of the ${\bar K}$ optical potential as
functions of the temperature for different densities.}
        \label{fig:temp7}
\end{figure}

In Fig. \ref{fig:temp7} we plot the real and imaginary parts of
$U_{\bar{K}}$ as functions of temperature for different densities.
It is interesting to observe that $U_{\bar{K}}$ depends very
weakly on temperature and stays attractive over the whole range of
temperatures explored. This is qualitatively different from the
results shown in Fig.~2 of Ref.~\cite{Schaffner}, where the
potential becomes repulsive at a finite temperature of $T=30 \
\rm{MeV}$ for $\rho=\rho_0$. We note that, although in that work a
self-consistent scheme was also applied, the only source of medium
effects in their Fig.~2 is the Pauli blocking of the nucleons in
the intermediate states. As noted in Ref.~\cite{Schaffner}, the
transition from attraction at $T=0$ to repulsion at finite $T$ is
in fact a consequence of the weakened Pauli blocking effects
associated to a thermal smearing of the Fermi surface, such that
eventually one is recovering the $\rho=0$ repulsive behavior. We
have checked that, under the same conditions, we obtain the same
qualitative behavior. Therefore, the results shown in
Fig.~\ref{fig:temp7} demonstrate that self-consistency effects
have a tremendous influence on the behavior of the ${\bar K}$
optical potential with temperature. In particular, for
temperatures as high as 70 MeV, the real part of $U_{\bar
K}(k_{\bar K}=0)$ at $\rho_0$ is very similar to that at $T=0$,
having lost only about 10 MeV of attraction. Hence, our results,
based on a fully self-consistent calculation, do not support the
claims that the ${\bar K}$ optical potential might be repulsive at
finite $T$. The attraction found here for the ${\bar K}$ optical
potential at finite $\rho$ and $T$, together with enhanced
in-medium $K^-$ production cross sections
\cite{Schaffner,Ohnishi97}, may help to explain the enhanced
$K^-/K^+$ ratio measured at GSI by the KaoS collaboration
\cite{kaos97}.

\begin{figure}[htb]
\centerline{
     \includegraphics[width=0.6\textwidth,angle=-90]{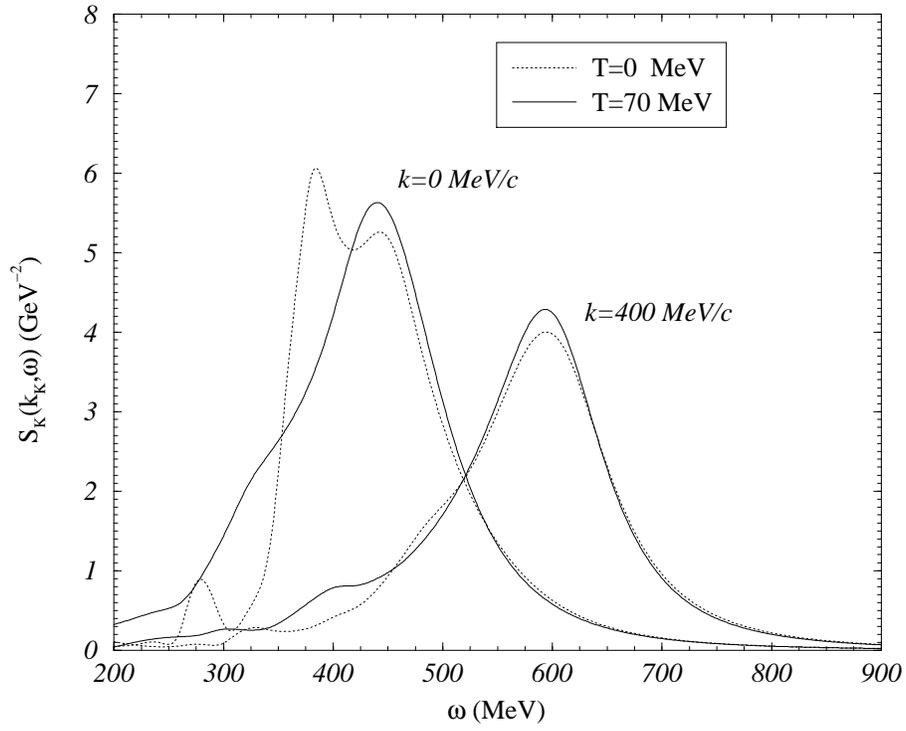}
}
      \caption{\small Spectral density of the antikaon at $\rho=0.17 \ \rm{fm^{-3}}$
      as a function of energy for antikaon momenta $k_{\bar{K}}=0$ and
      $k_{\bar{K}}=400 \ \rm{MeV/c}$, and for $T=0 \ \rm{MeV}$ and
      $T=70 \ \rm{MeV}$.}
        \label{fig:temp8}
\end{figure}

Finally, in Fig.~\ref{fig:temp8} we present the ${\bar K}$
spectral function as a function of energy for two momentum values,
$k_{\bar K}=0$ and 400 MeV/c, at $T=0$ (dotted lines) and $T=70$
MeV (solid lines). The structures observed on the left hand side
of the peaks, especially visible for the $T=0$ spectral functions,
are due to the excitation of hyperon-hole states with ${\bar K}$
quantum numbers that are present when the $L=1$ components of the
${\bar K}N$ interaction are incorporated. We note, however, that
the highest peak observed in the $T=0$ spectral function for
$k_{\bar K}=0$ corresponds to the enhancement observed in the
$I=0$ ${\bar K}N$ amplitude below the $\pi\Sigma$ threshold. The
inclusion of a finite $T$ washes most of these structures out, and
the resulting spectral functions show basically a single
pronounced peak.

\section{Conclusions}
\label{sec:conclusions}

We have studied the ${\bar K}N$ interaction in hot and dense
matter by extending to finite $T$ our previous $T=0$ model, which
is based on a self-consistent coupled-channel calculation taking,
as bare interaction, the meson-exchange potential of the J\"ulich
group.

Although more moderate than in the $T=0$ case, we have also found
at finite temperature that dressing the pions in the intermediate
$\pi Y$ loops has a strong influence on the ${\bar K}N$ amplitudes
and, consequently, on the ${\bar K}$ optical potential.

Partial waves higher than the $L=0$ component of the ${\bar K}N$
effective interaction contribute significantly to the antikaon
optical potential at finite temperature. The real part gains
attraction and the imaginary part becomes more absorptive. At a
momentum of 500 MeV/c, the contribution of the $L>0$ components is
as large as that of the $L=0$ ones.

We have found that self-consistency effects have a tremendous
influence on the behavior of the antikaon optical potential with
temperature. At normal saturation density, the optical potential
remains attractive for temperatures as large as 70 MeV, contrary
to what would be observed if only Pauli blocking medium effects
were considered, which would give a repulsive optical potential at
these high temperatures.

In general, temperature effects smear out the different
observables with respect to the $T=0$ case. For instance, the
antikaon spectral function at finite temperature shows much less
structure than that at $T=0$, reducing its shape basically to a
peak located at the quasiparticle energy.

The attractive potential found here for finite density, finite
temperature and finite momentum, i.e. for the typical scenario
found in heavy-ion collisions at GSI, is especially interesting to
understand the enhanced $K^-/K^+$ ratio measured by the KaoS
collaboration at GSI, together with other mechanisms that have
been already suggested in the literature, such as an enhanced
production of $K^-$ through $\pi Y$ collisions. Work along these
lines is in progress.

\section*{Acknowledgments}
We are very grateful to Dr.~J\"urgen Schaffner-Bielich and Dr. Wolfgang Cassing for useful
discussions. L.T. wishes to acknowledge the hospitality extended
to her at Brookhaven National Laboratory. This work is partially
supported by DGICYT project PB98-1247 and by the Generalitat de
Catalunya project 2000SGR00024. L.T. also wishes to acknowledge
support from the Ministerio de Educaci\'on y Cultura (Spain).

\newpage
\renewcommand{\theequation}{\Alph{section}.\arabic{equation}}
\setcounter{section}{1} \setcounter{equation}{0}
\section*{Appendix}
In this  appendix we show how to obtain the pion self-energy at
finite temperature. First, we will review the formalism at $T=0$.

When  the pion propagates in nuclear matter, it scatters with the
surrounding nucleons and the basic first order response of the
medium to the propagation  of the pion field is the excitation of
a particle-hole pair. In the region of intermediate energies
($T_{\pi}=0-300$ MeV), $\Delta-hole$ excitation also takes place
due to the excitation of internal degrees of freedom of the
nucleon. Therefore, to obtain the pion self-energy in symmetric
nuclear matter at $T=0$, both effects have to be taken into
account. We follow the nomenclature, parameters and approximations
mentioned in the appendix of Ref.~\cite{Oset90}. According to this
reference, the two contributions to the pion self-energy can be
written as

\begin{eqnarray}
\Pi(\vec{q},q^0 \,)= \left( \frac{f_N}{m_\pi} \right)^2 \vec{q}
\,^2 U(\vec{q},q^0 \,) \\ \nonumber U(\vec{q},q^0 \,)=
U_N(\vec{q},q^0 \,)+  U_{\Delta}(\vec{q},q^0 \,) \ ,
\end{eqnarray}
where $U(\vec{q},q^0 \,)$  is the sum of $1p-1h$ and $1\Delta-1h$
Lindhard functions, which at $T=0$ read

\begin{eqnarray}
&&U_N(\vec{q},q^0 \,)=  \label{u_nt0}
 \\ \nonumber \hspace*{-1.2cm}&&\nu_N \int
\frac{d^3k}{(2\pi)^3} \left[\frac {\theta(k_F-|\vec{k}|)
(1-\theta(k_F-|\vec{k}+\vec{q} \,|))}
{q^0+E_N(\vec{k})-E_N(\vec{k}+\vec{q} \,)+i\eta} +\frac
{\theta(k_F-|\vec{k}+\vec{q} \,|)
(1-\theta(k_F-|\vec{k}|))}{-q^0-E_N(\vec{k})+E_N(\vec{k}+\vec{q}
\,)+i\eta}\right]
\end{eqnarray}

\begin{eqnarray}
&U&_{\Delta}(\vec{q},q^0 \,)= \label{u_deltat0}
 \\ \nonumber && \hspace*{-0.4cm}\nu_\Delta \int \left[
\frac{d^3k}{(2\pi)^3}
\frac{\theta(k_F-|\vec{k}|)}{q^0+E_N(\vec{k})-
E_{\Delta}(\vec{k}+\vec{q} \,)+i \frac{\Gamma(\vec{q},q^0 \,)}{2}}
+\frac {\theta(k_F-|\vec{k}+\vec{q} \,|)}{-q^0+E_N(\vec{k}+\vec{q}
\,)-E_{\Delta}(\vec{k})+i\frac{\Gamma(\vec{q},-q^0 \,)}{2}}\right]
\ ,
\end{eqnarray}
where $\nu_N$ and $\nu_{\Delta}$ contain the sum  over
spin-isospin degrees of freedom in symmetric  nuclear matter. In
the case of $1\Delta-1h$, it also includes the decay constant
$f^*_{\Delta}$ fitted experimentally. Hence,
\begin{eqnarray}
\nu_N&=&4, \nonumber \\  \nu_{\Delta}&=&\frac{16}{9} \left(
\frac{f^*_{\Delta}} {f_N} \right)^2, \
\frac{f^*_{\Delta}}{f_N}=2.13 \ .
\end{eqnarray}
The quantity $\Gamma(\vec{q},\pm q^0 \,)$ is the $\Delta$ width,
which we take to depend only on the external variables of the
pion, as in Ref.~\cite{Oset90}.

The definitions of the Lindhard functions, Eqs.~(\ref{u_nt0}) and
(\ref{u_deltat0}), have been written explicitly here in order to
facilitate  the comparison with the finite $T$ case. On the other
hand, the analytical results can be found in the appendix of
Ref.~\cite{Oset90}.
%Both $W_N(\vec{q},q^0 \,)$ and $W_{\Delta}(\vec{q},q^0 \,)$ can be understood as the contribution of two many-body diagrams,
%direct and
%cross
%diagrams ,or  forward and backward propagating bubbles.
%(dibujo)
We also  include in the pion self-energy the coupling to $2p-2h$
excitations, following the phase-space approach of
Ref.~\cite{ROS94}.

Once $1p-1h$ ($2p-2h$) and $1\Delta-1h$  excitations are computed,
form-factors and short-range correlations are introduced. One pion
exchange provides a good description of the nucleon-nucleon
interaction for long distances, but this is not the only
ingredient. At shorter distances other effects must be considered,
such as correlated and uncorrelated two pion exchange and the
exchange of heavier mesons.
%It is mimic their contribution with an effective OPE interaction
%that can be translate into the following RPA sum,
%(dibujo)
%or a geometric progression.
The final expression for the pion self-energy reads
\begin{eqnarray}
\Pi(\vec{q},q^0 \,)&=&F(\vec{q},q^0 \,) \ \left( \frac{f_N}{m_\pi}
\right)^2  \vec{q} \, ^2 \frac{U(\vec{q},q^0
\,)}{1-(\frac{f_N}{m_\pi})^2 g'U(\vec{q},q^0 \,)}  \\ \nonumber
U(\vec{q},q^0 \,)&=&U_N(\vec{q},q^0 \,)+U_{\Delta}(\vec{q},q^0 \,)
\ ,
\end{eqnarray}
where $F(\vec{q},q^0 \,)$ is the form factor,
\begin{eqnarray}
F(\vec{q},q^0 \, )& = &\left( \frac{\Lambda^2-m_\pi^2}
{\Lambda^2-(q^{02}-\vec{q} \, ^2)} \right) ^2 \nonumber \\
\Lambda&=&1200\ {\rm MeV} \ ,
\end{eqnarray}
and $g^\prime$ the Landau-Migdal parameter taken from the
particle-hole interaction described in Ref.~\cite{OTW82}, which
includes $\pi$ and $\rho$ exchange modulated by the effect of
nuclear short-range correlations.

The pion self-energy also includes a s-wave piece
\begin{eqnarray}
\Pi(\rho)_{\rm{s-wave}}=-4 \pi \  (1+\frac{m_\pi}{m_N}) \
b_0 \ \rho,
\end{eqnarray}
with $b_0=\frac{-0.0285}{m_{\pi}}$, taken from the parameterization of
Ref.~\cite{SM83} , which is equivalent to the results of
Ref.~\cite{Meirav89}.

%The non-relativistic reduction for
%$\Pi(\vec{q},q^0 \,)$ that it is used for our calculations is defined as

%\begin{equation}
%U_{\pi}(\vec{q},q^0 \,)=\sqrt{\vec{q} \, ^2+m_{\pi}^2+\Pi_{\pi}(\vec{q},q^0 \,)}-\sqrt{\vec{q} \, ^2+m_{\pi}^2} \ .
%\label{eq:upi1}
%\end{equation}

\subsection{$1p-1h$ and $1\Delta-1h$ Lindhard functions at finite temperature}

The effect of temperature in the  pion self-energy comes from the
modification of  the Fermi sea. At a given temperature, nucleons
are distributed following the corresponding Fermi-distribution.
Then, $U(\vec{q},q^0 \,)$ transforms into
\begin{eqnarray}
&U&(\vec{q},q^0,T)=U_N(\vec{q},q^0,T)+U_{\Delta}(\vec{q},q^0,T)
\label{ay}
\end{eqnarray}
with
\begin{eqnarray}
\hspace*{-2cm} &&U_N(\vec{q},q^0,T)= \label{hey}
\\ \nonumber  && \nu_N
\int \left[ \frac{d^3k}{(2\pi)^3} \frac {n(\vec{k},T)
(1-n(\vec{k}+\vec{q},T))}{q^0+E_N(\vec{k})-E_N(\vec{k}+\vec{q}
\,)+i\eta} +\frac {n(\vec{k}+\vec{q},T)
(1-n(\vec{k},T))}{-q^0-E_N(\vec{k})+E_N(\vec{k}+\vec{q}
\,)+i\eta}\right]
\end{eqnarray}
and
\begin{eqnarray}
&U&_{\Delta}(\vec{q},q^0,T)= \label{hoy}
 \\ \nonumber
 &&\nu_{\Delta}
  \int \left[ \frac{d^3k}{(2\pi)^3}
\frac{n(\vec{k},T)}{q^0+E_N(\vec{k})- E_{\Delta}(\vec{k}+\vec{q}
\,)+i \frac{\Gamma(q^0,\vec{q} \,)}{2}} +\frac
{n(\vec{k}+\vec{q},T)}{-q^0+E_N(\vec{k}+\vec{q}
\,)-E_{\Delta}(\vec{k})+i\frac{\Gamma(\vec{q},-q^0 \,)}{2}}\right]
\ ,
\end{eqnarray}
where the step function $\theta$ for nucleons has been substituted
by the distribution $n(\vec{k},T)$ at the corresponding
temperature.

The imaginary part of $U_N(\vec{q},q^0,T)$  at finite temperature
can be obtained analytically

\begin{eqnarray}
{\rm Im} \, U_N(\vec{q},q^0,T)&=&{\rm Im} \, U_N^D(\vec{q},q^0,T)+
{\rm Im} \, U_N^C(\vec{q},q^0,T)= {\rm Im}
 \, U_N^D(\vec{q},q^0,T) \left( 1+{\rm e}^{\frac{-q^0}{T}} \right) \\
 {\rm Im} \,
U_N(\vec{q},q^0,T)&=&- \frac{3}{2} \pi \rho \frac{m_{\pi}^2
T}{k_F^3 |\vec{q} \,|} \ {\rm ln} \frac{1- n(p^+,T)}{1- n(p^-,T)}
\ {\rm cotanh}\left({\frac{q^0}{2T}}\right)  \label{imaun} \\ \nonumber
&&p^+=\frac{m_{\pi}}{|\vec{q}\,|} \left| q^0 +\frac{\vec{q} \,
^2}{2m_{\pi}} \right| , \ p^-=\frac{m_{\pi}}{|\vec{q}\,|} \left|
q^0 - \frac{\vec{q} \, ^2}{2m_{\pi}} \right|
\ , 
\end{eqnarray}
where ${\rm Im} \, U_N^D(\vec{q},q^0,T)$  and ${\rm Im} \,
U_N^C(\vec{q},q^0,T)$ are the imaginary parts of the direct and
crossed contributions (first and second terms on the r.h.s. of
Eq.~(\ref{hey}), respectively).

The real part of $U_N(\vec{q},q^0,T)$ is  obtained via the
dispersion relation
% The imaginary part is the sum of the direct and crossed
%terms. The dispersion relation reads,
\begin{eqnarray}
{\rm Re} \, U_N(\vec{q},q^0,T)= -\frac{1}{\pi}  P \int  d\omega' \
\frac{ {\rm Im} \, U_N^D(\omega',\vec{q},T)}{q^0-\omega'}
-\frac{1}{\pi} P \int  d\omega' \ \frac{{\rm Im} \,
U_N^C(\omega',\vec{q},T)}{-q^0+\omega'} \ , \label{realun}
\end{eqnarray}

or,  in a more compact form,
\begin{eqnarray}
{\rm Re} \, U_N(\vec{q},q^0,T)= -\frac{1}{\pi}  P \int  d\omega' \
\frac{{\rm Im} \, U_N(\omega',\vec{q},T) \ {\rm
tanh}\left(\frac{\omega'}{2T}\right)}{q^0-\omega'} \ .
\label{realdos}
\end{eqnarray}

Eqs. (\ref{imaun}) and (\ref{realdos})  define the nucleon
Lindhard function at finite $T$.

As mentioned in the appendix of Ref.~\cite{Oset90}, the
$1\Delta-1h$ Lindhard function at $T=0$, $U_{\Delta}(\vec{q},q^0
\,)$, can be derived analytically by neglecting the difference of
$\vec{k}^2$ terms in

\begin{eqnarray}
E_N(\vec{k})-E_{\Delta}(\vec{k} \pm \vec{q} \,)
=\frac{\vec{k}^2}{2m_N}-\frac{\vec{k}^2}{2m_{\Delta}}-\frac{\vec{q}
\, ^2}{2m_{\Delta}} \mp
\frac{\vec{k}\vec{q}}{m_{\Delta}}+m_N-m_{\Delta} \ ,
\end{eqnarray}
and assuming that the $\Delta$ width only depends  on the external
variables. We perform the same approximations at finite $T$.
Redefining $\vec{k}+\vec{q} \rightarrow -\vec{k}$ in the second
term on the r.h.s of Eq.~(\ref{hoy}), such that it becomes the
same as the first one but changing $q^0 \rightarrow -q^0$, we
finally arrive at

\begin{eqnarray}
U_{\Delta}(\vec{q},q^0,T)&=&\frac{2}{3} \left( \frac{f^*_{\Delta}}
{f_N} \right)^2 \frac{m_{\Delta} \rho}{q k_F^3} \int dk \,  k  \,
n(k,T) \, \left[ \ln \left( \frac{z^++1}{z^+-1} \right) +\ln
\left( \frac{z^-+1}{z^--1} \right) \right] \\ \nonumber
z^{\pm}&=&\frac{m_{\Delta}}{q \ k} \left( \pm q^0-\frac{\vec{q} \,
^2}{2m_{\Delta}}-(m_{\Delta}-m_N)+i\frac{\Gamma(\pm q^0,\vec{q}
\,)}{2} \right) \ .
\end{eqnarray}

\newpage
\renewcommand{\theequation}{\Alph{section}.\arabic{equation}}

%\setcounter{section}{2}
%\setcounter{equation}{0}

%%%%%%%%%%%%% FIGURE 1 %%%%%%%%%%%%%%%%%%%%%%%%%%%%%%%%%%
%%%%%%%%%%%%%%%%%%%%%%%%%%%%%%%%%%%%%%%%%%%%%%%%%%%%%%%%%%%
%%%%%%%%%%%%% FIGURE 2 %%%%%%%%%%%%%%%%%%%%%%%%%%%%%%%%%%%
%%%%%%%%%%%%%%%%%%%%%%%%%%%%%%%%%%%%%%%%%%%%%%%%%%%%%%%%%%%
%%%%%%%%%%%%% FIGURE 3 %%%%%%%%%%%%%%%%%%%%%%%%%%%%%%%%%%
%%%%%%%%%%%%%%%%%%%%%%%%%%%%%%%%%%%%%%%%%%%%%%%%%%%%%%%%%%%
%%%%%%%%%%%%% FIGURE 4 %%%%%%%%%%%%%%%%%%%%%%%%%%%%%%%%%%
%%%%%%%%%%%%%%%%%%%%%%%%%%%%%%%%%%%%%%%%%%%%%%%%%%%%%%%%%%%
%%%%%%%%%%%%% FIGURE 5 %%%%%%%%%%%%%%%%%%%%%%%%%%%%%%%%%%
%%%%%%%%%%%%%%%%%%%%%%%%%%%%%%%%%%%%%%%%%%%%%%%%%%%%%%%%%%%
%%%%%%%%%%%%% FIGURE 6 %%%%%%%%%%%%%%%%%%%%%%%%%%%%%%%%%%
%%%%%%%%%%%%%%%%%%%%%%%%%%%%%%%%%%%%%%%%%%%%%%%%%%%%%%%%%%%
%%%%%%%%%%%%% FIGURE 7 %%%%%%%%%%%%%%%%%%%%%%%%%%%%%%%%%%
%%%%%%%%%%%%%%%%%%%%%%%%%%%%%%%%%%%%%%%%%%%%%%%%%%%%%%%%%%%
%%%%%%%%%%%%% FIGURE 8 %%%%%%%%%%%%%%%%%%%%%%%%%%%%%%%%%%
%%%%%%%%%%%%%%%%%%%%%%%%%%%%%%%%%%%%%%%%%%%%%%%%%%%%%%%%%%%

\end{document}